\documentclass[a4paper,12pt]{article}
\usepackage[utf8]{inputenc}
\usepackage{cite}
\usepackage{amssymb}
\usepackage{amsmath}
\usepackage{amsfonts} % for mathbb{1} indicator
\usepackage{amsthm}

\theoremstyle{definition}
\newtheorem{definition}{Definition}[section]
\theoremstyle{plain}
\newtheorem{theorem}{Theorem}[section]

\providecommand{\keywords}[1]{\textbf{Keywords: } #1}

\title{Deliberative Democracy \\
with Dilutive Voting Power Sharing
}
\author{Dimitrios Karoukis}

\begin{document}

\maketitle

\begin{abstract}

We present a deliberation model where a group of individuals with heterogeneous preferences
iteratively forms expert committees whose members are tasked with the updating of
an exogenously given status quo change proposal.
% We present a collective decision-making model where
% a status quo change is exogenously proposed to a group of individuals who then
% iteratively update it through an elected, dynamic, deliberative committee of their experts.
Every individual holds some initial voting power
that is represented by a finite amount of indivisible units with some underlying value.
Iterations happen in three stages.
In the first stage,
everyone decides which units to keep for themselves and where to distribute the rest.
With every ownership mutation,
a unit's underlying value diminishes by some exogenously given amount.
In the second stage, the deliberative committee
is formed by the individuals with the most accumulated voting power.
These experts can author corrections to the proposal which are proportional to their
accumulated power.
In the third stage,
if an individual outside of the committee disagrees with a correction,
she can vote against it with their remaining voting power.
A correction is discarded if more than half of the total voting power
outside of the committee is against it.
If either the committee or the proposal remain unchanged for two consecutive
iterations, the process stops. We show that this will happen in finite time.

\end{abstract}


\section{Introduction}

% Representative democracies whose legislative bodies have
% constant members over extended periods of time and across different subjects
% may fail to represent society effectively due to possible opinion divides
% between their members and their voters.
% If there is lack of communication between the legislators and
% the society then this divide can become significant and, in the case of
% decision-making, it can become detrimental for the common good.
%
% In representative democracies, opinion divides between those with
% legislative power and the electorate may deteriorate the quality of
% if there is no interaction between the two during any stage
% between the drafting and the final vote.
In a representative democracy,
meaningful and frequent interaction between the citizens and their representatives
during the various stages of the legislative process, from the drafting
of a law to its final vote,
safeguards the quality of the democracy's institutions from deterioration.
However, if the members of the legislative bodies remain
constant over extended periods of time,
the risk of them lacking expertise as individuals in the subject of some law
that they are called to decide upon during their term increases overtime.
In order for representation to work properly,
there needs to exist a mechanism that lets society pick its
representatives according to its perception of their expertise on each subject in question.
This mechanism could also allow for updating of these representatives during the
process in order to protect the electorate from corrupt individuals.

Suppose that a status quo change proposal has been drafted exogenously
and the affected group of individuals wants to manipulate its text according to their
preferences.
In this model, the group forms a deliberative committee consisting of its elected experts,
whose function is to make corrections to the proposal prior to it becoming permanent.
This committee could be seen as a type of parliament.
The committee members are assumed to be iteratively updated
through a modified version of \textit{liquid democracy}
that is used as a mechanism to help the group dynamically identify its
experts for the subject in question.
% We call the process \textit{liquid deliberation} and its objective is to
% update the laws proposed by the legislators with the opinions of the voters.

Liquid democracy \cite{blum2016},
or more accurately \textit{liquid voting},
is an extension of proxy voting \cite{miller1969} where a
voter $A$ can either vote for themselves or delegate
their voting power to some other voter $B$.
Voter $B$ can decide to vote with
their increased voting power or delegate it all to another voter $C$
and so on, making these delegations transitive.
Liquid voting was designed to help in situations where
a voter realises that they do not possess sufficient knowledge to make
an informed decision and they feel more comfortable
assigning their democratic right to another voter who
they deem more capable of making an informed decision.

In our approach,
we use a generalisation of liquid voting which was briefly discussed
in \cite{zhang2021}.
In this context,
voter $A$ can delegate \textit{shares} of their voting power
to one or more voters who can also transitively delegate shares of
their voting power to others and so on.
The voters may delegate shares that add up to less than $100\%$
of their voting power and keep the rest for themselves.
It could be used as a \textit{vote hedging} strategy
in the sense that it may reduce the risk induced by
incomplete information about the knowledge possessed
by the receiving voters on the subject in question.
We represent these shares as indivisible \textit{voting power units}
whose underlying value suffers
a predefined amount of dilution with each change in ownership.

There are numerous other variations of liquid voting in the literature.
Many of them exist because a shortcoming of pure liquid voting has been
shown in \cite{kling2015} to be the tendency of some voters to become
\textit{super-delegates} by accumulating a large part of the total voting
power in the society, while the majority of voters never receive any
delegations.
This accumulation of voting power has been shown in \cite{kahng2021} to be
harmful for social welfare, even if these super-delegates are competent
individuals.

A remedy that was suggested in \cite{golz2018} is to let voters
choose more than one candidates to potentially receive their voting power and
let a centralized mechanism assign the voting power to the right candidates
according to the \textit{law of communicating vessels}.
This law of physics suggests that a homogeneous fluid in a set of connected
containers will always balance out to the same level regardless of the shape
of the containers.
In our context, we can think of the homogeneous fluid as the voting power and
the containers as the candidate voters.
The centralized mechanism that follows this law will ensure that
no voter among the candidates will
end up with excess voting power relative to the other candidates.

% Mention \cite{valsangiacomo2021} and \cite{civicracy2016}.

The inspiration for the deliberative committee of experts and the updating of the proposal
is drawn from from the DeGroot model of
social influence as described in \cite{degroot1974}, \cite{chatterjee1977}
and \cite{golubjackson2010}.
In this model, weighted averaging is used in order to
update the belief that an agent has about the state of the world
depending on the weight that they assign to the beliefs of their
acquaintances. In our model, instead of
subjective probability distributions over the state of the world,
status quo change proposals are represented by vectors in the $[0, 1]^s$ set,
with $s$ being the number of subjects that will be affected
and $[0, 1]$ the range of values that each subject can attain.
Every individual in the society has an \textit{opinion} regarding
the status quo change that depends on her information set and is
represented by a preference over
every possible status quo change that can affect these $s$ subjects.

The choices of every individual in the society regarding
the distribution of their voting power in every iteration, which happens in three stages,
are as follows.
If they take no action in the first stage of the initial iteration $t=0$,
they have chosen to keep all of their voting power units to themselves.
If they want to distribute them,
they announce the units to be transferred and their recipients.
In the first stage of every subsequent iteration,
inaction means that they will stick with their previous distribution.
In the second stage of every period there is no choice regarding
the use of voting power to be made.
In the third stage, a vote needs to be cast only if the individual disagrees with a correction that was proposed by a deliberative committee member.
If at any $t > 0$ they announce a change in their voting power distribution,
the units that are going to be redistributed will lose a percentage of their value.

The members of the dynamic deliberative committee of experts
are the individuals with the most accumulated voting power
at the second stage of every iteration.
Their accumulated power is represented by the units that they kept for themselves
and the dilution that these units have suffered until that point.
Each member can update the proposal's text to be closer to
their opinion by replacing or adding text whose percentage of the
corpus is less than or equal to the percentage of
the total available voting power of the committee
that they hold at $t$.
If more than half of the voting power of the society
outside the committee at $t$ is against a specific correction,
it will be discarded.

% We also restrict the number of individuals who propose status quo
% changes in the beginning to be less than a
% third of the number of experts in the deliberative committee
% because they can also be members of the committee.

% It will also protect the system from malicious voters
% who perform unending, cheap and inconsequential changes in delegation.
% A final assumption is that at each $t > 0$,
% a voter will either change their voting power
% distribution or they will abstain and transfer the entirety of their
% voting power to the legislators for that period.
% We will refer to this as the \textit{quicksand} assumption.
% Since we are in the liquid voting
% paradigm, they can always change their choice in a subsequent period.

% At some time $T < \infty$,
% and all individuals in the society participate in a referendum
% in order to decide whether to make the updated proposal permanent or
% not.
% The deliberations stops
If either the committee members and their voting power or the proposal's text
remain unchanged for two consecutive time periods,
the deliberation stops.
We show that at least one of the two conditions will be satisfied in finite time,
so there is no need for predefined restrictions on
the number of iterations that need to take place before the process stops.
The assumption that voting power units depreciate in value with every mutation in their
``chain of custody'' ensures this assertion.
It also ensures that an individuals with many directly transferred units has
better chances to be in the expert committee compared to a peer with more
indirectly transferred units.

\section{Model}

Suppose society $N$ of $n < \infty$ individuals and
deliberative committee of experts at iteration $t<\infty$
consisting of $k < n$ members, denoted by $K_t \subset N$.

\begin{definition}\label{def:statusquo}
The \textit{status quo} is represented by the vector $$\ell_{-1} \in [0, 1]^s,$$
where $s$ indicates the subjects of interest and
$[0, 1]$ the range of values that each subject can attain.
The \textit{initial} status quo \textit{change proposal}
is exogenously given and represented by $\ell_0 \in [0, 1]^s$
such that $\ell_0 \neq \ell_{-1}$.
Every subsequent proposal at $t$ is represented by $\ell_{t} \in [0, 1]^s$.
\end{definition}
%
% group $G$ of $|G| = g$ individuals proposing the initial status quo
% change that may or may not be part of $N$,
% but if $G \subset N$ then $g < \overline{k}/3$.

The voting power of an individual is represented by
\textit{voting power units}, which are elements in the set
$
U \subseteq [0, 1] \times N^\omega,
$
% $$U \equiv \left\{[0, 1], \{i\}_{i \in N}^v | 1 \leq v \leq n\right\}$$
with $1 \leq \omega \leq n$ and $|U| = u < \infty$.

\begin{definition}\label{def:valuechain}
At some iteration $t < \infty$,
a voting power unit with \textit{index} $id \in [0, u]$
$$
u_{id}(t) \in U
$$
has \textit{value} represented by a real number
$$
u_{id}^{value}(t) \in [0, 1]
$$
and
\textit{chain of custody} represented by
a vector
$$
u_{id}^{chain}(t) = \{i\}_{i \in N}^\omega \in N^\omega
$$
of size $\omega \leq n$,
ordered from their original owner to the latest.
\end{definition}

The last element of the chain of custody of a unit at $t < \infty$ is represented by
$u_{id,\omega}^{chain}(t)$.
The original owner is immutable, but every other owner in the chain of custody
can be mutated by owners who come before them.

\begin{definition}\label{def:votingpower}
The \textit{voting power} of an individual $i \in N$ at iteration $t < \infty$
is measured by the sum of all units' values
whose last owner in the chain of custody is $i$, or
\begin{equation}\label{eq:power}
P_i(t) = \sum_{id = 0}^u \left\lbrace u_{id}^{value}(t) | u_{id,\omega}^{chain}(t) = i \right\rbrace.
\end{equation}
\end{definition}

At $t = 0$,
every $i \in N$ is assigned a finite number of units with value equal to one and chain
of custody only including the original owner, or
$$u_{id}^{value}(0) = 1 \text{ and } u_{id}^{chain}(0) = i.$$
The conceptually simplest case would be for every individual to be assigned $100$ units of
voting power at $t=0$ but the model can also accommodate scenarios where
individuals start the process with different amounts.

In every iteration $t$, there are three stages to the deliberative process.
In the first stage, each individual $i \in N$ decides whether to keep their
voting power units to themselves or to delegate
some or all of them to one or more individuals.
The receivers can transitively delegate some or all of
these units to other individuals etc.
In the second stage,
the distribution of voting power stops and
the top $k$ individuals with respect to their total voting power
are the experts chosen by the society to form
the deliberative committee at $t$.
% In the third stage,
% those outside of the committee can vote negatively on corrections to the proposal.
The third stage of the process is where the proposal is updated
but it will be described in detail in section \ref{sec:minting}.
% Assume for simplicity that \textit{every} individual $i \in N$ has
% \textit{initial} voting power $p^i_0 = 1$ such that the
% \textit{total voting power of the society at any $t$} is
%
% \begin{equation}\label{total power of society}
% \sum_{i \in N} p^i_t = n.
% \end{equation}

% Assuming a fully connected societal graph,
% meaning that every citizen $i$ can delegate
% their voting power to any other citizen $j$, we have
% a left-stochastic voting power matrix of the form
% \begin{equation}\label{eq:power matrix}
%     W_t =
%     \begin{bmatrix}
%         w_t^1(1) & \hdots & w_t^1(n) \\
%         \vdots     & \ddots & \vdots     \\
%         w_t^n(1) & \hdots & w_t^n(n)
%     \end{bmatrix},
% \end{equation}
% with every row summing up to $1$,
% representing the distribution of voting power.
% In the case where everybody has kept their voting power
% to themselves, $W_t$ becomes a $n\times n$ identity matrix.

If, at any $t < \infty$, $i \in N$ changes their voting power distribution,
% or $w_{t}^i(j) \neq w_{t-1}^i(j)$ for any $j \in N$,
the distributed units' values are diminished by an exogenously given \textit{dilution}
factor $c \in (0, 1)$.
% that is multiplied by the number of times she has already changed her
% voting power distribution $\lambda_{t-1}^i$ plus one,
% with $\lambda_0^i = 0$ for all $i$.

\begin{definition}
The distribution of voting power units from some $i \in N$ to some $j \in N \setminus \{i\}$
\textit{mutates} the set $U$ by diluting the value of the units by an exogenously given
factor of $c \in (0, 1)$ \textit{and}
by making $j$ the last element of the chain of custody right after $i$.
In functional form, the distribution is a map $T_{i, j}: U \rightarrow [0,1] \times N^\omega$
such that

\begin{equation}\label{eq:distribution}
\begin{split}
T_{i, j}^{value}\left(u_{id}(t)\right) &= (1 - c) u_{id}^{value}(t), \\
T_{i, j, \omega - 1}^{chain}\left(u_{id}(t)\right) &= u_{id, \omega}^{chain}(t) = i \text{ and } \\
T_{i, j, \omega}^{chain}\left(u_{id}(t)\right) &= j.
\end{split}
\end{equation}

\end{definition}

The experts that comprise the committee in the second stage of period
$t$ are the top $k$ individuals in the society
with respect to their voting power (\ref{eq:power}).
To break possible ties in voting power amount,
we compare the various-order \textit{appeals} of the competing experts,
meaning that the winner of the tie is the one
to whom the most voting power was transferred to \textit{directly} or
\textit{indirectly} but at the same order.

\begin{definition}\label{def:appeal}
The \textit{first-order appeal} of $i \in N$ at time $t < \infty$ is the \textit{sum} of
voting power units that were assigned to her \textit{directly} by $j \in N\setminus \{i\}$,
or whose $u_{id}^{chain}(t) = [j, i]$.
\end{definition}

\begin{definition}\label{def:vappeal}
The $\omega$-\textit{th-order appeal} of $i \in N$ at time $t < \infty$ is the sum of units
that were assigned to her from their $(\omega - 1)$'th owner, or whose
$\left|u_{id}^{chain}(t)\right| = \omega$ and $u_{id,\omega}^{chain}(t) = i$.
\end{definition}

If no individual in the society has transferred any voting power units to another
by the end of the first stage,
the process stops and the status quo change proposal remains unchanged.
If less than $k$ individuals in the society can successfully break the tie
then the deliberative committee consists of only those individuals
who hold more than the tie-inducing voting power amount.

% we draw inspiration from proof-of-work algorithms used in
% blockchain technology and we can assume that it is broken
% by making all the competing prospective members simultaneously guess
% a number as close as possible to an exogenously and randomly chosen number
% in $\mathbb{R}$ that is revealed to all after the guess.
% The prospective member whose guess is closest to that number joins $k_t$.
% This will be repeated until all ties are broken and we assume
% that it will finish at some $\tilde{T} < \infty$ with probability 1.

\section{Proposal Minting}\label{sec:minting}

At the end of the second stage of each time period
there are at \textit{most} $k$ candidate corrections
to the proposal $\ell_t$,
since the members of the committee are the only ones that can author
corrections but they are not obliged to exercise their right.
The size of each correction relative to the corpus of $\ell_t$ is
proportional to the percentage of the \textit{total} voting power of the
committee that each $\sum_{i \in k_t} P_i(t)$ that each expert holds.

The updating of the corpus happens in the third stage of each time
period and it happens in two steps.
First, the members of the committee propose corrections and the
citizens that are not in the committee vote for or against them.
Second, amendments to some corrections are made and the citizens
outside of the committee vote on them.

Assume that there is a function
$$f: [0, 1]^s \times [0, 1]^s \rightarrow [0,1]$$
denoting the percentage of the corpus of $\ell_t$
that needs to change due to a correction brought forward
by a member of the committee.

\begin{definition}\label{def:dell}
A correction to the corpus of $\ell_t$ brought forward
by a member of the committee $i \in k_t$ is denoted by
$d\ell_{t+1}^i$ and it needs to satisfy the constraint

\begin{equation}\label{eq:dell}
f\left(d\ell_{t+1}^i, \ell_t\right) \leq \frac{P_i(t)}{\sum_{j \in k_t}P_j(t)}.
\end{equation}
\end{definition}

% \begin{definition}\label{def:dell}
% A member of the committee $i \in k_t$ brings forward at $t$
% a correction to the proposal $\ell_t$ that is of the form
%
% \begin{equation}\label{eq:dell}
% d\ell_{t+1}^i = \frac{1}{p_t^N}
% \left\lbrace
%     \ell^i p_t^i(i) + \ell_t\left[p_t^N - p_t^i(i)\right]
% \right\rbrace.
% \end{equation}
% \end{definition}

A member of the committee $i \in k_t$
can also build on the correction of another member of the committee
$j \in k_t \setminus \{i\}$ and
their correction also needs to satisfy (\ref{eq:dell}).
% \begin{equation}\label{eq:dell_j}
% d\ell_{t+1}^i = \frac{1}{p_t^N}
% \left\lbrace
%     \ell^i p_t^i(i) + d\ell_{t+1}^j\left[p_t^N - p_t^i(i)\right]
% \right\rbrace.
% \end{equation}

If an individual $i \notin k_t$ with $P_i(t) \neq 0$
does not agree with a correction to the proposal at $t$,
they can vote against it.
If more than half of the total voting power of the rest of the
society is against a correction then it is blocked.
In order to require minimum engagement by those outside of $k_t$,
we assume that their vote is positive by default
unless they explicitly disagree with the correction.

\begin{definition}\label{def:opinion}
Each individual $i \in N$ has an \textit{opinion} regarding status quo changes,
which is described by a total order\footnote{
A total order is a binary relation $\succsim$ on a set $X$ which satisfies,
for any $a, b, c \in X$, the following:
\begin{enumerate}
    \item $a \succsim a$ (reflexivity)
    \item if $a \succsim b$ and $b \succsim a$ then $a = b$ (antisymmetry)
    \item if $a \succsim b$ and $b \succsim c$ then $a \succsim c$ (transitivity)
    \item $a \succsim b$ or $b \succsim a$ (totality).
\end{enumerate}
} $\succsim_i$ over the set $[0, 1]^s$ and
an optimal proposal $\ell^i \in [0,1]^s$ such that
$
\ell^i \succsim_i \ell \text{ for any } \ell \in [0, 1]^s.
$
% For the sake of simplicity,
% we will assume that this opinion does not change overtime.
\end{definition}

% \begin{definition}\label{def:p_t^N}
% The total voting power of the society that is not in the deliberative
% committee is denoted by $p_t^N$ and it is of the form
% \begin{equation}\label{eq:p_t^N}
% p_t^N = \sum_{i \in N \setminus \{k_t\}} p_t^i(i).
% \end{equation}
% \end{definition}

\begin{definition}\label{def:vote}
The \textit{vote} of an individual $i \in N \setminus k_t$ on a $j \in k_t$
committee member's correction $d\ell_{t+1}^j$ is a function
$v: [0,1]^s \rightarrow \mathbb{R}_+$ that is equal to $P_i(t)$ if the corrected proposal is
preferable to the status quo and to the latest state of the proposal and zero otherwise.
This function is of the form
\begin{equation}\label{eq:vote}
v_j(d\ell_{t+1}^i) = P_j(t) *
\mathbf{1}\{d\ell_{t+1}^i \succsim_j \ell_t \succsim_j \ell_{-1}\}.
\end{equation}

\end{definition}

The \textit{condition} that a correction $d\ell_{t+1}^i$ needs to satisfy in order
to become part of $\ell_{t+1}$ is that more than half of all the voting power outside of the
committee is \textit{favorable} to it, or
\begin{equation}\label{eq:condition}
\begin{split}
&\sum_{j \in N \setminus k_t} v_j(d\ell_{t+1}^i)
>
\frac{1}{2}\sum_{j \in N \setminus k_t} P_j(t).
\end{split}
\end{equation}

If two conflicting corrections $d\ell_{t+1}^i$ and $d\ell_{t+1}^j$
of some $i, j \in k_t$ both get accepted,
$i$ and $j$ need to generate a new correction at the second
step of the process, which
% will be of the form
% $$
% d\ell_{t+1}^{ij} =
% \frac{d\ell_{t+1}^i p_t^i(i) + d\ell_{t+1}^j p_t^j(j)}
% {p_t^i(i) + p_t^j(j)}
% $$
% and it
needs to pass (\ref{eq:condition}) or it will not be
included in $\ell_{t+1}$.

If $d\ell_{t+1}^j$ builds upon $d\ell_{t+1}^i$ for any $i, j \in k_t$
and either $j$'s correction or both corrections get rejected we do not
have any ambiguous outcomes.
If both corrections are accepted then the richest one,
in the sense that it builds upon the most
corrections, is added to $\ell_{t+1}$.
If one or more corrections upon which a correction $d\ell_{t+1}^j$
is built are voted down then $j$ can propose a different correction,
whose inclusion in $\ell_{t+1}$ is subject to
(\ref{eq:condition}).

\section{Results}

% Each committee $k_t$ holds an internal vote in each period $t > 0$
% where if more than $1/2$ of its members vote in favour of stopping
% the process then it stops and the referendum is held.
The process is assumed to stop naturally if no corrections have been made to
the proposal between $t-1$ and $t$.
This can mean that either
no member of $k_t$ brought forward a correction
to the proposal at $t$ or that every correction that was brought
forward was voted down by the rest of the society.
The process is also assumed to stop if the resulting
committee's members and their voting power remain unchanged
between $t-1$ and $t$.
A final assumption that was mentioned earlier is that
if at any time $t < \infty$ it is true that
$P_i(t) = P_j(t)$ for all $i, j \in N$ then
the deliberation stops.
% We also define a hard limit $\overline{T} < \infty$ such that
% $T < \overline{T}$.

\begin{theorem}\label{thm:finality}
There exists some $T < \infty$ when the deliberative process stops.
\end{theorem}
\begin{proof}
By assumption, the process stops if the proposal remains unchanged for two
consecutive iterations or if the members of the committee and their
voting power remain unchanged for two consecutive iterations.
If the experts and their voting power change then some $i \in N$
has redistributed their voting power units,
which means that these units' value has diluted by $c \in (0, 1)$.
As the process continues,
more and more citizens will have diluted voting power units.
Since the number of individuals in the society and
the number of total available voting power units are finite,
and $c \in (0, 1)$,
eventually all citizens will hold zero voting power and
this will happen in finite time.
By the assumptions, this will stop the deliberation.
\end{proof}

\section{Future research}

In the future, work on economic-theoretic questions would shed light on matters
such as the existence of Nash equilibrium deliberative committees,
their stability and their representation of citizens' preferences.
We will then be able to answer
questions such as the optimal size $k$ of the deliberative
committee with respect to the size of the society
or the optimal penalty $c$ etc.
We could also introduce the possibility to purchase
voting power at some point in the process,
maybe with some quadratic cost as in \cite{weyl2012}.

An assumption that can be relaxed is the fully connected societal
graph.
In practice, people would not transfer their
voting power units to anyone in the society that is outside of their \textit{social
network}, unless they were a "celebrity".
The proximity of citizens in the social graph could be a restricting
factor for the number of other citizens that a citizen can assign
their voting power to.
%
% We could also allow for citizens to have a dynamic $\ell^i$ at $t+1$
% that may depend on the $\ell^i$ of the previous period,
% the opinion of their representative and the proposed law at $t$.
% We could also add a random noise
% $\epsilon \sim \mathcal{N}(0,\sigma^2)$
% to their opinion in every period.
%
% If the system was to be implemented in a society,
% the state could also offer extra initial voting power to its citizens as an
% incentive for them to contribute in some positive way in public life.

\section{Conclusion}

We introduced a form of governance that can bridge
the gap between the will of the legislators and the voters.
If the legislators remain invariant over time periods, we proposed
the formation of a sort of dynamic "parliament",
a deliberative committee of experts,
that negotiates status quo change proposals in order
to ensure that the opinions of the voters are taken into account
during the legislative process.
By making this committee dynamic,
we also give the voters the power to
mitigate possible adverse effects of corruption of its members.
By getting constant signals on the sentiment about the corrections of
the proposal, the chosen experts can make informed decisions about
what type of changes to the status quo do the voters want implemented.
It is our hope that the proposed system will help improve
representation of peoples' preferences in collective decision making
settings.

\bibliography{liquid_deliberation}
\bibliographystyle{plain}
\end{document}